# Droplet Epitaxy of Semiconductor Nanostructures for Quantum Photonic Devices

Massimo Gurioli, *University of Firenze, Firenze, Italy*
Zhiming Wang *Institute of Fundamental and Frontier Sciences, University of Electronic Science and Technology of China, Chengdu 610054, P.R.China*
Armando Rastelli, *J. Kepler University, Linz, Austria*
Takashi Kuroda, *National Institute of Materials Science, Tsukuba, Japan*
Stefano Sanguinetti*, *University of Milano Bicocca, Milano, Italy*
* corresponding author: stefano.sanguinetti@unimib.it

**Abstract**

The long dreamed 'quantum internet' would consist of a network of quantum nodes (solid-state or atomic systems) linked by flying qubits, naturally based on photons, travelling over long distances at the speed of light, with negligible decoherence. A key component is a light source, able to provide single or entangled photon pairs. Among the different platforms, semiconductor quantum dots are very attractive, as they can be integrated with other photonic and electronic components in miniaturized chips. In the early 1990's two approaches were developed to synthetize self-assembled epitaxial semiconductor quantum dots (QDs), or 'artificial atoms', namely the Stranski-Krastanov (SK) and the droplet epitaxy (DE) method. Because of its robustness and simplicity, the SK method became the workhorse to achieve several breakthroughs in both fundamental and technological areas. The need for specific emission wavelengths or structural and optical properties has nevertheless motivated further research on the DE method and its more recent development, the local-droplet-etching (LDE), as complementary routes to obtain high-quality semiconductor nanostructures. The recent reports on the generation of highly entangled photon pairs, combined with good photon indistinguishability, suggest that DE and LDE QDs may complement (and sometime even outperform) conventional SK InGaAs QDs as quantum emitters. We present here a critical survey of the state of the art of DE and LDE, highlighting the advantages and weaknesses, the obtained achievements and the still open challenges, in view of applications in quantum communication and technology.

Among the research and technology areas which are leading to the "second quantum revolution", quantum communication is arguably the most advanced. Commercial quantum cryptographic systems are already available, offering enhanced security by using keys encoded in the states of quantum systems (photons)[1]. The next step could be a large quantum network, over which photons will transport quantum information, to specific quantum nodes where it would be processed or stored.[2] For the implementation of a reliable quantum internet, a key ingredient is the development of quantum light source integrated into photonic chips. So far two quantum states of light have played a major role: indistinguishable single photons and entangled photon pairs. Indeed, entangled photon pairs are necessary for many applications[3], such as the teleportation[4,5] of quantum states between arbitrarily distant locations, which, in turn, is a key resource to alleviate unavoidable transmission losses in the network[6] as well as for device-independent quantum cryptography.[7–9] While there are several competing systems as efficient quantum emitters of single photons (e.g. atoms, organic molecules, colour centres, quantum dots, carbon nanotubes)[10] as well as several approaches for integrating them into photonic chips, semiconductor quantum dots (QDs) obtained by epitaxial growth are the only solid-state sources, which have so far demonstrated the capability of producing triggered entangled photon pairs.[11,12]

Semiconductor-based QDs are heterostructures, made of crystals composed of more than one coherently grown material, whose electronic states in the conduction and valence bands can be precisely tailored. The spatial variation of the semiconductor composition over nanometric scales leads to a spatial modulation of the band gap, which is used to control the carrier motion and eventually to obtain localization of the electrons and holes in one or more spatial directions. Whenever the electronic wavefunction size becomes comparable or smaller than the De Broglie wavelength of carriers (around few tens of nanometers for common semiconductors), quantum confinement effects become pronounced. Then the optical and electronic properties of the heterostructure depend on the carrier localization size. Whenever the carriers quantum confinement occurs in all the three spatial dimensions, the quantum nanostructures (having compact shapes: lenses, pyramids, cones etc.) are usually referred to as quantum dots (QDs). In QDs only discrete energy states are allowed and their electronic density of states (DOS) is therefore a series of delta functions. The possibility to use QDs for concentrating the DOS in a reduced energy range proved important for a large variety of fundamental topics and device applications. QDs have been systematically explored to improve "classical" optoelectronic devices e.g. semiconductor laser with extremely low threshold[13], while they are considered promising solid-state building blocks for emerging quantum technologies: they can act as deterministic sources of single photons [14] and quantum entangled photon pairs.[15] Quantum light sources are required in several applications (see refs [16–18]), such as quantum communication and photonic-based quantum simulation and computing.

In the 1980's, the dream of atomic-like density of states in semiconductor devices led the scientific community to try at first top-down approaches based on lithography and etching of quantum wells (QWs) or local band gap modulation [19], with the results being rather poor due to defects induced during the post-growth processing. In the same years, chemical bottom-up approach leaded to the vast field of colloidal semiconductor QDs.[20] Finally, in the early 1990's, two bottom-up epitaxial approaches emerged, leading to self-assembled QDs of III-V semiconductor materials: i) the SK growth mode, which takes advantage of strain for 3D island formation on top of a thin wetting layer (WL) of the same material of the nanostructure [21,22] and ii) the Droplet Epitaxy (DE) method, exploiting the controlled crystallization of metal nano-droplets into III-V semiconductors [23]. Both methods, derived from standard epitaxial growth modes, lead to coherent self-assembly of defect-free and optically active III-V QDs. The robustness and simplicity of SK growth led to its large dominance, with great achievements in both fundamental and technological areas[24–26].
The SK growth occurs almost exclusively during lattice mismatched heteroepitaxy. In III-V systems, he archetypal example of SK QDs is represented by InGaAs QDs grown on GaAs(001) substrates and overgrown with GaAs[26]; with InAs having a lattice constant ~7% larger than GaAs. In a simplified picture the growth evolves as follows: initially InAs (or InGaAs) wets the GaAs surface and the growth proceeds layer by layer, giving origin to the WL. Above a critical thickness of the WL, the lattice-matching conditions between the epitaxial layer and the substrate increase the strain energy stored in the epilayer, causing a change of the surface morphology from flat to 3D (island mode). In spite of the huge success of SK in the research community, there are several unavoidable constraints of this class of QDs: i) the presence of a 2D WL interconnecting the QDs, ii) the residual strain and associated built-in piezoelectric fields, iii) the limited range of island shapes - which are mostly determined by the energetically favored facets and QD size [22,27] – iv) the strain-enhanced intermixing occurring during growth and capping - which leads to complex composition profiles[28] v) the difficulty of controlling separately the QD structure and surface density, vi) the limited combinations of materials composition which translates in a limited range of emission frequencies accessible with SK QDs, and vii) the limited number of crystal orientations of the substrate. All these effects have not prevented SK QDs to set the state-of-the-art for QD-based single photon sources,[10,17] but they limit the possible material chart, complicate the growth on high Miller index substrates and usually require post growth tuning to obtain "entanglement-ready"

QDs.[29] Driven by the high flexibility in the nanostructure shape design and the unique possibility to produce QDs out of lattice-matched materials, research on DE has continued over the years and is now showing its potential to widen the range of applications of epitaxial QDs as ideal quantum light sources for quantum communication and photonic-based quantum simulations.

## 2. Growth

DE QD and SK QD share many common features in terms of electronic states and optical properties. Whereas SK is superior in terms of ease of implementation, DE offers additional advantages. In the following description of the DE procedures we will concentrate our discussion on the prototypical DE QDs made of GaAs embedded in an AlGaAs barrier[30], but similar processes – possibly in different temperature ranges due to element specific binding energies and diffusivity – can be applied to a variety of different group III -V combinations.[23,31–35]

**2.1: Metal droplet assembly**:

The first step in the growth process is the deposition of the Ga metallic atoms on an AlGaAs layer (Figure 1a). A metal-stabilized surface is usually required for the droplets to form.[36] Afterwards, either by excess metal deposition or As desorption, the additional deposition of group-III metal adatoms leads to the spontaneous formation of nanometer-scale droplets, as water droplets on a glass window, irrespective of the lattice mismatch; the formation of Ga droplets on the surface follows the Volmer–Weber growth mode.[37,38]

At this stage, it is possible to control the droplet density (between $10^3$ to 0.1 droplets $\mu m^2$, Figure 1b)[39,40] and size (between 10 and 30 nm) [39], by changing the substrate temperature between 150 and 500 °C (which affects the droplet density) or by the choice of metal atom coverage (total amount of metal deposited per unit surface area)[41]. The size distribution of DE QDs can be as low as 10%, which is comparable to the size dispersion of SK QDs[42]. Much broader size distributions (up to 25%, as required in case of optical amplifiers) can be obtained by high metal fluxes and/or low deposition temperatures[43].

**2.2: Droplet Crystallization**:

The subsequent step is the droplet crystallization by the annealing in group V element atmosphere, during which each single metal droplet becomes the starting point of one nanostructure, whose shape and topology can be tailored by controlling the process[44]. Here we distinguish DE, based on the use of the metal contained in the droplet to fabricate a QD, from the more recent local droplet etching (LDE), exploiting the initially formed droplets to etch nanotemplates on the substrate for the subsequent growth of the QDs. The two methods differ in the substrate temperatures and As flux used during the annealing process, as explained in the following sections.

**2.2.1: Droplet Epitaxy**:

DE fabrication of 3D GaAs nanostructures is controlled by two main processes[38]:
1. the As incorporation in the metallic droplet and the GaAs crystallization at the liquid-solid interface between the droplet and the substrate (red arrow in the Figure 1c);
2. the As adsorption on the surface surrounding the droplet, which changes the surface termination and provides the driving force for the diffusion of Ga out of the droplet by capillary forces (blue arrow in the Figure 1c).

The As atoms impinging on the droplet during the crystallization step are dissolved into the liquid metal droplet; the droplet is then crystallized by nucleation at the interface with the surface[45]. On (001) oriented substrates, the crystalline GaAs that is formed at this stage appears as a ring at the perimeter of the droplet. The ring then thickens due to subsequent direct As incorporation into the droplet. On (111) substrates instead, the droplet crystallizes in form of hexagonal or triangular base islands, due to the intrinsic surface symmetry.[46] At the same time, Ga adatom diffusion and Ga-As bond formation around the droplet lead to the accumulation of GaAs within the diffusion length of Ga adatoms from the droplet edge.

Temperature, arsenic pressure and surface orientation govern the balance between process 1 and 2 in DE. Their competition during As exposure permits to crystallize the metal contained in the droplet in a large variety of shapes, which can be controlled by the growth parameters (substrate orientation, temperature and As flux) through the modulation of the Ga migration on the surface. Qualitatively, Ga migration is prevented at low temperature, because of the thermal activation of this process, and hindered at higher As pressure[47]. Ga diffusion lengths vary between 1 and 60 nm when the temperature is changed within the typical DE range (150 °C – 350 °C) at the As pressure of $10^{-5}$ Torr on (001) substrates[47]. In the cases dominated by growth at the liquid-solid interface within the droplet (process 1) compact island morphologies are observed (Figure 1d, top). On the opposite, when diffusion from the droplet edge controls the Ga crystallization dynamics (process 2), hollow morphologies, like disks and single rings, can be formed (Figure 1d, bottom). Higher temperatures lead to growth modes close to the formation of thin GaAs layers around the droplet site[48] and substrate local etching (see below). In between these two extremes, more complex morphologies, like double rings and molecules, can be obtained[44,49,50]. It is worth noticing that (111)A oriented substrates allow crystallization of the Ga droplets into QD under As supply even at relatively high temperatures (around 500 °C). The prevalence of the growth at the liquid-solid interface is attributed to the short surface residence time of $As_4$ which reduces surface reactivity.[51]

As mentioned above, the range of QD morphologies and sizes attainable via SK is limited, due to constrains on elastic, interface, and surface energies (even if capping and annealing can be used to tune the size and shape of the QDs).[27] DE QDs maintain their morphology upon capping (Figure 2a)[52], thanks to the limited (below 1 nm) interdiffusion that takes place at the QD-barrier interface (still, a large and controlled interdiffusion at the QD-barrier interface is possible through low temperature capping and post-growth annealing[53]). The size and aspect ratio of DE QDs can be directly tuned in a broad range during the island formation (Figure 2b).
Decoupling the dot formation from strain relaxation offers a larger material chart with respect to SK, which may be important for the improvement of QD-based devices, like single photon emitters, detectors or solar cells[54–56]. DE QDs have been reported in several semiconductor materials, like InSb/CdTe[57], In(Ga)As/GaAs[58,59], GaAs/Si[32], GaSb/GaAs[33], GaN/AlGaN[34] and tensile-strained GaP/GaAs [35]. As far as surface orientation is concerned, highly symmetric QDs were obtained by DE on (111) substrates [60,61]. The (111) substrate orientation is appealing for its high in-plane symmetry, but problematic for SK growth, since on (111) the relaxation of a strained III-V semiconductor epilayer immediately proceeds through the nucleation of misfit dislocation at the interface rather than through the formation of coherent 3D islands [62].

A (sometime undesired) feature of the SK QDs is the unavoidable presence of the WL at the base of the dots, which generates bidimensional electronic states interconnecting the dots. This affects the QD optical properties and carrier kinetics[63]. The WL is also detrimental in device performances (eg lasing threshold), since it represents a channel for carrier escape out of the QDs [64]. Even in strain-free GaAs DE-QDs grown on (001) substrates, a thin layer interconnecting them can be formed during droplet deposition. However, the WL can be entirely avoided by controlling the surface reconstruction and composition before the droplet formation[63]. On the other hand, owing to the substantial independence of the growth substrate, DE QDs can be then realized on top of a quantum well of the desired thickness and composition[63] (Figure 2c).

Finally, the shape control by DE on (001) leads to form quantum rings in a variety of configurations[44,49], based on the interplay between diffusion and crystallization effects during the annealing of the droplets in As atmosphere.[49,65,66] In addition, the growth can be tailored in real time, by applying a series of short As pulses, and tuning As fluxes and substrate temperatures. The crystallization of a controlled fraction of the Ga reservoir in each growth step permits the combination of quantum dots, rings and disks in a single nanostructure[44,50].

**2.2.2: Local Droplet Etching:**

One drawback of DE, which has hindered for many years its widespread use, is the rather low temperature used on (001) substrates during the crystallization of the metal droplets (essential to avoid bidimensional growth), which leads to poor material and optical quality. It can nevertheless be overcome by in-situ and ex-situ annealing[53,67,68] or by the use of (111)A substrates.[69] Still, subtle effects related to the presence of defects, such as spectral diffusion of the QD exciton states[70] (which can be considered detrimental for quantum applications), are usually larger in DE QDs than in SK QDs; this limit has been overcame by a recent evolution of the DE technique, the local-droplet-etching (LDE)[71].

The LDE process can be divided into four major steps (Figures 3a-d)[72,73]: a) Formation of metal nanodroplets, as in DE; b) Local etching; c) Removal and recrystallization of residual metal; d) QD formation by hole filling. The local etching (of the III-V semiconductor underneath the droplet, usually AlGaAs) occurs when the nanodroplets are annealed under low group-V flux and relatively high substrate temperatures[74]. The fine control of all the growth parameters determines the nanohole depth, shape, and lateral extension.[75]. When the As concentration exceeds the solubility in the metal (~0.1% in Ga at 580 °C and <0.1% at lower temperatures), arsenic forms covalent bonds with metal atoms at the droplet edges, leading to a nanohole surrounded by a crystalline wall, followed by the appearance of a circular ridge around it and a thin bidimensional layer (Figure e-f)[76]. Extending the annealing/etching time until all metal has recrystallized can remove all the residual metal from the bottom of the nanohole[73]. Since nanohole etching takes place under epitaxial growth conditions, it leads to defect- and impurity-free "nanopatterning", which is not the case for *ex-situ* etching methods. The self-assembled nanoholes have been used as a template for various types of structures, including low density QDs[77] and coupled lateral QD pairs[78] by controlling the sequence of materials during the hole filling.

The main application of LDE is the fabrication of high temperature strain-free QDs [79–81]. The average QD emission energy and thickness of the epitaxial layer interconnecting the QDs be easily controlled by tuning the amount of GaAs used during the hole filling[82] (Figure 3g). Depending on the growth and overgrowth conditions, surface mounds can be observed after capping, which can serve as intrinsic markers to precisely locate the buried QDs and position additional nanostructures on them, such as plasmonic antennas[83]. In addition to providing excellent optical properties, QDs fabricated by nanohole filling can exhibit high symmetry, imposed by the nanohole, a useful trait for the generation of entangled photon pairs)[80].

**3 Quantum Photonics with DE QDs**

DE allows to expand the chart of possible QD heterostructures; tuning the emission energy of the exciton recombination by about 1 eV to achieve single photon emission from telecom to visible wavelengths has been reported[34, 62, 60,84] (Figure 4a). Light-emitting-diodes (LEDs), lasers, detectors, solar cells based on DE QDs have been reported as well[85], but the best reported performances still lag behind those for SK QDs-based devices. Nevertheless, the peculiarities of DE and the lifting of several SK constraints is expected to lead to advantages in term of higher quantum efficiency in optoelectronic devices[85]. Here we will focus on the recent achievements of DE QDs on two aspects in the emerging field of quantum applications: emission of single photons and entangled photon pairs. Table 1 provides a summary of their key features along those of SK, in terms of growth control and optical properties.

## 3.1 Electronic and optical properties

The possibility to control 3D nanostructure shape opens fascinating scenarios for fundamental studies. Various phenomena are expected to appear for complex nanostructures such as rings and disks in conjunction with QDs. The DE ring structures show the presence of a single quantum state extending through the whole ring [86], thus permitting the observation of Aharonov-Bohm oscillations by spectroscopic measurements, which imply the formation of strongly correlated exciton pairs (Wigner molecules) formed in a non-simply connected topological nanostructure[87]. DE complexes, including QDs and rings[44,49,50], offer multiple quantum interference trajectories for charge carriers, switchable via external perturbations with internal carrier dynamics determined by topology-dependent selection rules [88].

The optical selection rules in QDs made of common direct-bandgap semiconductors are largely determined by the hole fundamental state. For widely studied InAs or InGaAs QDs, the strong compressive strain splits the heavy hole (hh) and the light hole (lh) levels, and the lowest-energy exciton has a dominant hh character. For strain-free GaAs/AlGaAs QDs, hh and lh splitting is relatively small and the character of the hole ground state is easily controllable by application of external stress[89]. Strong hh-lh mixing enables to systematically tune anisotropic optical response, thanks to precise control in QD geometry, which can be achieved via in-situ annealing of GaAs/AlGaAs DE QDs[90]. For asymmetrically shaped dots, hh-lh mixing leads to a brightening of the nominally dark exciton and also to an anomalous anticrossing of the bright exciton when anisotropic stress is applied [91]. Emergence of optically-forbidden lines in magneto luminescence spectra allows to clarify the impact of $C_{3v}$ symmetric carrier confinement on strong valence band mixing (Figure 4b) [92]. In addition, large hh and lh mixing results in efficient hole-nuclei coupling, which is normally weak compared with electron-nuclei coupling for strained QDs [93]. Enhanced electron-nuclei, as well as hole-nuclei coupling leads to strong dynamic nuclear polarization even at zero magnetic field.[94]

## 3.2 Single photon emitters

Single QDs are currently among the most promising solid-state sources of single photons on demand[10]. Following the pioneering works around 2000[14,95], nearly ideal performance in terms of single photon purity, brightness and indistinguishability have been recently reported for resonantly-driven InGaAs SK QDs embedded in micropillar cavities[96,97]. The engineered photonic environment boosts both the brightness of the sources and the indistinguishability of the emitted photons due to the Purcell effect, which accelerates the spontaneous emission rate and mitigates the effects associated to charge noise and – in narrow-band cavities – also of phonons. For a recent review including the requirements for ideal single photon sources and their applications, we refer the reader to Ref.[17]

GaAs QDs obtained by DE and LDE operate at wavelengths (700 nm-800 nm) matching the high-sensitivity spectral range of state-of-the-art silicon-based single-photon detectors, making them attractive for free-space quantum optics experiments[98]. Their emission can be tuned to match the D1 and D2 transitions of $^{87}$Rb[81,99,100], providing the opportunity to interface the emitted photons with quantum memories based on warm atomic vapors (see Figure 4c)[101]. In turn, these systems would allow storing the state of the photons at nodes of a quantum network.

In order to target the indistinguishability, the single photon emission should be Fourier limited and the fast radiative lifetime of DE-QDs (in the range of 250 ps) is an advantage in this respect. Indeed QDs almost at the Fourier limit (not exploiting Purcell effect) have been reported both in DE and LDE [69,102] and very recently it was reported that charged excitons confined in GaAs QDs grown by LDE may exhibit Fourier-limited spectral linewidths under resonant cw excitation[99]. In the same experiment, resonantly scattered light was used to perform high-resolution spectroscopy on a Rb vapour. In term of purity of the $|1\rangle$ Fock state (i.e. zero value of second order intensity correlation function $g_2(0)$), similar to SK QDs, LDE-GaAs QDs display almost negligible probability of emitting more than one photon per excitation cycle (Figure 4d). By using LDE QDs embedded in a planar cavity, two-photon excitation, and superconducting low-noise single-photon detectors, the lowest probability of multiphoton generation ($g_2(0) \sim 7 \times 10^{-5}$) for any quantum emitter to date (including single atoms and ions) has been achieved[103]. The purity and indistinguishability of single photons emitted by single LDE-GaAs QDs under pulsed, resonant two-photon excitation of the biexciton (XX) state has also been investigated[104]. The degree of indistinguishability of photons (i.e. the visibility of the Hong-Ou-Mandel interference dip), when emitted upon excitation cycles separated by 2 ns, was found to reach values up to ~90% without the need for spectral filtering.

Ideally not only the photons subsequently emitted by a single QD, but also photons emitted by different emitters should be indistinguishable for the implementation of a quantum network, which is largely more demanding. A typical shortcoming of solid-state quantum emitters is the fact that not all QDs in an ensemble have the same high optical quality nor the same spectrum; this drawback is common to both for SK QDs and for DE QDs. This limitation is however not a major block since two-photon interference between single photons emitted by remote GaAs LDE QDs was recently demonstrated by using a phonon-assisted two-photon excitation scheme (Figure 4f)[102]. The visibility of the two-photon interference, obtained by tuning the emission of one of the QDs via piezoelectric-induced strain, reached levels of ~0.5 without any spectral filtering, which is the highest value reported so far for QDs (see Figure 4g). This result is partly due to the short (~250 ps) radiative lifetime of confined excitons, compared to ~1 ns for typical SK-InGaAs QDs. This relevant feature of DE QDs is not yet totally explained by quantitative theoretical models and it is tentatively ascribed to the enhanced oscillator strength in QDs larger than the exciton Bohr radius in GaAs[105,106] (the base diameter of the GaAs-filled nanoholes is ~40 nm, to be compared to a Bohr radius of ~12 nm). Even if further studies are needed for a full understanding of the point, this is a clear advantage of DE-QDs since it partially relaxes the need of resorting to the Purcell effect in optical microcavities to increase the radiative emission rate of excitons, a route commonly followed with InGaAs SK dots, where the radiative lifetime is in the range of ns[96].

For practical single-photon emitting devices, it would be of the utmost relevance to replace the optical excitation with electrical injection. Compared to other quantum emitters (organic molecules, NV centers, atoms, etc.), QDs are ideally suited for this purpose, because they can be easily incorporated in LED structures. The first single-photon LED based on an InGaAs QD was reported already in 2002[107] and good performances in terms of brightness and photon indistinguishability were achieved later[108–110]. In the case of DE and LDE QDs, single-QD LEDs with tunable emission wavelength have been recently reported[111] (Figure 4h).
Still, we have to acknowledge that this is a topic which needs further efforts, since the performance of such electrically-driven single photon sources has not yet reached the level of those operated under resonant optical injection, especially with respect to photon indistinguishability. This can be attributed to the time jitter introduced by the random capture of carriers by the QD and additional charge noise introduced by the doped layers near the QDs.

Another step toward the practical implementation of quantum devices is the coupling of QDs with photonic structures, and particularly with photonic crystal microcavities[112], which becomes optimum upon the spatial superposition of the QD states and the confined photon modes. The most effective strategy is the fabrication of the cavity after locating the QD by microscopy techniques.[113,114] But, for the implementation of QDs in a fully designable and complex photonic circuit, the control of the QD nucleation position at the nanometer scale is essential. The most successful approach is the self-limited growth of QDs into inverted pyramidal holes etched in a GaAs substrate.[115,116] Very good control of the SK QD nucleation site has been also obtained via patterning the substrate with nanohole arrays.[117–119] The integration of DE and LDE GaAs QDs in photonic structures relies so far on the same strategies used to control SK QD nucleation sites[120,121] but the processing and etching of different materials remains challenging. Bragg mirrors for GaAs emitters have been extensively studied[17], leading to a very easy route for integrating DE QDs in pillars and planar photonic cavities. On the contrary in the field of photonic crystals on membrane, the need of an $Al_xGa_{1-x}As$ (usually x=0.3) membrane for the GaAs QDs may raise problems in the residual roughness of the alloy membrane (likely limiting the Q factor of the photonic modes) and on the parasitic surface recombination at the air pores inside the AlGaAs membrane. Let's mention that even in case of the unconventional QDs on (111), the integration in photonic structures[122] and the electrical injection does not pose novel challenge.[123,124] Still other issues are present, such as the necessity of a high growth rate to fabricate high Q planar microcavities by DBRs.

For applications not requiring high degree of indistinguishability, such as for short-distance quantum communication, QD-LEDs could be useful in term of suitable wavelength, but operation above the typical liquid-He temperature, as well as integration on silicon substrates are crucial and open points. First steps have been taken on effective DE QDs integration on Si and Ge substrates exhibiting photon antibunching up to 80 K[125]. Higher temperature is so far a far dream both for DE and SK QDs. On one side emission broadens due to the contribution of phonon side-bands, which deteriorates photon indistinguishability. This is common to all high T quantum emitters and it could be optimized by integration in photonic resonators.[126,127] More complex for QDs is to avoid multiphoton emission for the exciton an biexciton cascade, that are almost degenerate transitions at room temperature. This feature which is an advantage of QDs at low T, since it allows for entangled photon generation, eventually ends up as a stumbling block for room temperature operation as quantum emitters, likely to be tackle by using more polar semiconductors with strong carrier confinement and large spectral separation between different excitonic species are required for possible high-temperature operation, such as nitride alloys. Single photon operation from QDs at temperatures ranging from liquid nitrogen to room temperature have been reported for these materials.[128–131] Record high-temperature single photon emission has been demonstrated by GaN/AlN site-controlled GaN QDs in nanowires[128]. Emission of such dots however occurs in the UV spectral range, which is currently above the optimum sensitivity range of single-photon detectors. In this respect, InGaN/GaN QDs self-assembled by DE[132], showed, for the first time, single photon emission up to the temperature of 80 K at around 440 nm (blue).[130] Nevertheless, approaches with nitride heterostructures face common challenges, such as strong background signals and significant spectral diffusion of the QD emission. The large built-in electric fields in the nitride QDs due to piezoelectric and spontaneous polarization is a disadvantage in terms of repetition rate and indistinguishability of the generated photons due to the induced large electron-hole separation in the QD, which makes the polarized excitons particularly sensitive to charge noise.[133] The reduction or elimination of the built-in field can be obtained by using nitride QDs assembled by DE on non-polar substrates[34] or even in lattice matched conditions.

**3.3 Entangled photon pairs emitters**
An important feature of QDs is their capability to generate, at low T, polarization-entangled photon pairs by using the cascade decay of a biexciton state $|XX\rangle$ to the crystal ground state $|0\rangle$ through the intermediate bright exciton $|X\rangle$ states[12]. This makes QDs different from other quantum emitters, such as organic molecules, carbon nanotubes[134] and color centers (even if dyads can support biexciton[135] no direct demonstration of entanglement has been so far reported). Another widespread solid state route to entangled photons are the nonlinear parametric down-converters (PDC)[136] which are intrinsically based on laser pumping with Poissonian (non-deterministic) photon statistics, even in the limit of low pumping.

QDs, on the other hand, are capable of generating entangled photon pairs "on demand"[137,138]. By denoting the photon emitted by the $|XX\rangle \rightarrow |X\rangle$ decay as XX and the one emitted by the $|X\rangle \rightarrow |0\rangle$ decay as X, the time-dependent two-photon state generated by the cascade can be written as [139]: $\frac{1}{\sqrt{2}}\left[(XX_H X_H) + e^{\frac{i\delta t}{\hbar}}(XX_V X_V)\right]$, where H and V denote "horizontally" and "vertically" polarized light (parallel to the transition dipoles of the QD exciton) and $\delta$ the energy splitting (fine-structure-splitting, FSS, due to the electron-hole interaction) between the two bright excitonic states of a heavy-hole exciton. The FSS is produced by the electron-hole exchange in presence of an anisotropic potential in the QD growth plane and has been subject to numerous studies[87,140–142]. To achieve high degree of entanglement without resorting to ultrafast detectors (eg. capable of discriminating the temporal oscillations introduced by the time-varying phase), the FSS must be small compared to the spectral bandwidth of the emitted photons (~3 μeV for a lifetime of 250 ps).

As already discussed, for conventional SK InGaAs QDs grown on GaAs(001) substrates, controlling the FSS is challenging because of the interplay of anisotropic shape, composition gradients, strain and piezoelectricity. The experiments on entangled photon generation have thus mostly relied on careful preselection of QDs which, for statistical reasons, presented sufficiently small FSS[143–145]. In fact, only very recently it has been possible to restore the degeneracy of the excitonic states for arbitrary QDs by introducing suitable external perturbations[146,147].

The study of strain-free GaAs DE QDs, excluding strain and piezoelectric effects, allowed to obtain a clear correlation between the observed FSS and structural data[148]. Figures 5a reports the FSS variation for (001) GaAs DE QD as a function of the exciton recombination energy (i.e. the QD size). In large DE QDs the shape elongation on the [1,-1,0] axis is the main responsible for the exciton FSS. The geometrical asymmetry and thus the FSS can be strongly suppressed by reducing the QD size (a residual FSS in the range of tens of μeV has been attributed to randomly distributed charge centers[70]). Trigonally symmetric (111) substrates for the self-assembly of QDs[60] can provide an additional route to control the FSS: for the (111) plane equivalent directions appear with respect to every 120º rotation, thus removing in-plane anisotropy in the QD structure. Data reported in Figure 5a demonstrates a suppression of the FSS independent of QD size for GaAs/AlGaAs QDs formed on Ga-rich (111)A substrates. The same concept was used to obtain InAs/InAlAs QDs on InP(111)A with almost zero FSS and emission wavelength in the infrared telecom window[61].
A clear correlation between QD shape/size and FSS was experimentally observed in LDE-grown GaAs/AlGaAs QDs on GaAs(001) substrates [80] (Figure 5b) and reproduced by model calculations with no adjustable parameters[82]. By optimizing the nanohole shape on AlGaAs, QDs with average FSS of 4 μeV (comparable to the radiative linewidth of the emission lines) were obtained[104]. LDE permits to improve also the overall QD ensemble homogeneity, with a much sharper FSS distribution when compared to the InAs SK QDs (Figure 5c).

DE InAs QDs emitting in the high transparency window of conventional optical fibers (1.55 μm) and embedded in planar cavities on conventional InP(001) substrates, were found to display a mean FSS four times smaller than SK-InAs QDs[149]; such characteristics have been exploited for demonstrating a quantum LED at 1.55 mm[150] paving the way for the implementation of entangled-photon-pair sources compatible with existing fiber optics.

The small mean value of FSS in GaAs QDs obtained by DE [84] and LDE [29,104], has led to entanglement fidelities to the Bell state as high as 86(±2)% and clear violation of Bell's inequalities without spectral or temporal filtering[84] using photon pairs. Even higher values up to 94(±1)% were obtained for a LDE QD [104] in spite of a finite FSS value of 1.2 μeV (Figure 5d-e).

While available theoretical models predict an upper limit for the degree of entanglement achievable with InGaAs QDs [151], near-unity fidelities can be expected in strain-free GaAs QDs (Figure 5e). LDE QDs with FSS tuned below the detection limit via modetate externally-induced strains exhibit a record fidelity of 97.8(±0.5)% under resonant two-photon excitation.[152]

The self-assembly of QDs by DE on intrinsically symmetric (111) surfaces permits raising the yield of entanglement-ready photon sources to values as high as 95%.[69] This is due to the low values of fine structure splitting and radiative lifetime induced by the fundamental droplet crystallization step at a significantly high temperature. Despite the limitations exposed in section 3.2, this technology is compatible with integration in photonic crystals for enhanced light extraction,[96,122,153] and constitutes an ideal candidate for semiconductor-based sources of entangled photons.

**4.0 Summary and outlook**

The prospect of using epitaxial QDs as semiconductor-based hardware for optoelectronic and quantum optics applications drove the scientific community to a tremendous effort to develop high quality QDs and understand their physical properties. Many challenges on the control of material quality, homogeneity, density, stacking, strain, piezoelectric fields, size, shape, nucleation site, have been addressed. Given its ease of fabrication, SK growth dominated the scientific and technological development of epitaxial QD nanoscience. However, despite the remarkable progress in SK flexibility and versatility, several constraints remain. In this front, DE can extend the range of achievable QD design degrees of freedom, namely materials, size, shape and density, while allowing for precise tuning of the symmetry, aspect ratio, topology, and, finally, high-quality GaAs QDs in AlGaAs barriers, a material combination for which SK cannot be applied.

These morphological aspects are reflected in peculiar optical properties such as large range of the exciton emission, no FSS, integration on Si, short exciton lifetime, to mention few of the most relevant ones. These DE-QDs excitonic features start to be considered of the utmost importance in the field of quantum emitters for quantum communication. In this emerging topic LDE QDs have been used to achieve the best performance to date in terms of background-free single-photon generation [103] and feature emission in the spectral range of the D1 and D2 transitions of Rb-vapours. On the other hand, the strongly reduced in-plane asymmetry of DE and LDE QDs [29,69,80,104] have recently enabled the demonstration of the highest degree of entanglement for photon pairs emitted by QDs.[152]

Integration of DE and LDE QDs in photonic structures, even if still in its infancy should enhance and optimize the forthcoming quantum devices (along the lines of SK QDs integration[96,153]). Whether the achievable performance in terms of photon indistinguishability will be sufficient to scale-up the "quantum hardware" to multiple sources is still an open question, not only for DE and LDE QDs but for all solid-state quantum emitters. Photon indistinguishability is affected by interactions of confined excitons with phonons as well as charge and nuclear fluctuations.

While the wavelength range accessible with GaAs QDs is potentially suitable for satellite-based quantum communication, the integration with classical fiber-based telecommunication technologies is essential but requires quantum emitters at 1550 nm. This region is not accessible with most organic molecules or color centers, but it has been already reached by DE QDs grown on InP substrates[61,150] We expect DE to be likely superior to SK in this aspect. DE QDs may lead to single photons as well as entangled photon pairs without resorting to brightness-degrading temporal filtering, a result not reached so far at 1550 nm.

As the degree of control on the QD properties following self-assembly (SK, DE and LDE QDs) will remain limited due to unavoidable stochastic processes occurring during growth, ideal devices will require the capability of controlling the QD properties after growth. The use of external perturbations, such as strain and electric fields, appears at present the most promising route to overcome this limitation, as recently demonstrated for InGaAs QDs and also LDE QDs[147].

Finally, the development of room temperature (RT) quantum emitters is of the utmost importance. Organic molecules, color centers and nitride QDs are suitable candidate; they all however fail to reach telecom wavelengths. The development of epitaxial QDs efficiently emitting at room temperature is limited by the carrier escape form the QDs to the barrier, but heterostructures with higher band gap mismatch seems to be the key. The development of hybrid system on InP-based QDs with GaN barriers seems to be the natural solution for efficient quantum emitters in the telecom bands at room temperature, but of course other unexpected solutions may arise on the way. In conclusions, we predict that the additional degrees of freedom offered by DE could increase the chances of success of QDs as useful building block for the quantum communication revolution, even if only relevant developments in the next future are needed to fulfil our prediction and hope.

Quantum Dots. *Phys. Rev. Lett.* **107,** 166604 (2011).

| Feature | SK | DE | LDE |
|---|---|---|---|
| Control of nanostructure shape during growth | Moderate | High | Moderate |
| Control of composition profile during growth | Poor | High | High |
| Control of wetting layer thickness | Poor | High | High |
| Control of structural properties during capping | Moderate | Moderate | High |
| Growth on (111)-oriented substrates | Difficult [154] | Yes | - |
| Lattice matched growth | No | Yes | Yes |
| Minimum mean excitonic fine-structure splitting for entangled photons | ~6 μeV (only after post-growth annealing [155]) | 4.5±3.1 μeV[68] [GaAs/AlGaAs(111)A] | 4.3 μeV [80] [GaAs/AlGaAs(001)] |
| Maximum entanglement fidelity | 0.8 | 0.86 | 0.94 |

**Table I:** Comparison between key features of SK, DE and LDE QDs respect to the feasibility of nanostructure growth control and to relevant optical properties for quantum applications.

**Figures**

**Figure 1:** (a) Droplet Epitaxy schematics: In an MBE chamber, group III metals are deposited at first on the surface (upper panel), leading to the formation of droplets. After this step (lower panel), the metal source is closed, and group V flux is supplied to the surface to crystallize the droplets into 3D nanostructures. (b) Examples of high density (n = 3 X $10^3$ $\mu m^{-2}$) and low density (n = 4 $\mu m^{-2}$) QDs obtained on AlGaAs(001) substrates by varying the substrate temperature between 150°C and 350°C during metal deposition. (c) Schematics of the DE QD growth mechanism during group V (typically As) supply. Process 1 (red arrows): As direct incorporation in the metallic droplet gives rise to GaAs crystallization at the liquid-solid interface, starting from the triple line. Process 2 (blue arrows): Owing to As adsorption on the surface surrounding the droplet, surface termination is changed and diffusion of Ga out of the droplet by capillary forces take place. Coming from the droplets edge, Ga atoms can migrate covering a mean diffusion length before being incorporated into the GaAs crystal. Blue dots indicate Ga atoms. As atoms are indicated by yellow dots. (d): AFM image (right) and atomistic scheme (left) of a DE QD (top) and a hollow DE nanostructure (bottom). The DE QD was obtained crystallizing the droplet T=200 °C and As flux ≈ $10^{-5}$ Torr, thus favouring As direct incorporation into the droplet during the crystallization step. The hollow nanostructure was obtained crystallizing the droplet T=350 °C and As flux ≈ $10^{-7}$ Torr on (001) substrates, where process 2 (out diffusion of metal from the droplet and incorporation in to the bulk) dominates.

**Figure 2:** (a) 40 nm X 34 nm topographic image of a typical GaAs/AlGaAs QD by cross sectional STM (courtesy of Paul Koenraad, Technical University of Eindhoven) [156] (b) Aspect ratio tuning range dependence on temperature in DE QDs. The QDs are crystallized, from the same droplet configuration. Inset, DE QDs crystallized at 150 °C (left) and 250 °C (right). The QD heights are 40 nm and 15 nm, respectively. (c) Cross-section TEM images. Top: QDs without wetting layer; Bottom: QDs with 3 ML wetting layer (Ref. [157])

**Figure 3:** (a-d) Sketch of the LDE nanohole fabrication process. The deposited droplet partially etches the substrate underneath due to As diffusion from substrate into droplet and partial recrystallization at the droplet rim up to equilibrium concentration (b). After As supply the etching of the nanohole starts. The completed nanohole (c) also includes a crystallized droplet rim and a two dimensional layer (whose composition depends strongly on the initial droplet material) surrounding the hole. The LDE QD can be obtained by filling the nanohole with a suitable material and subsequently capping it. (f-h) Illustration (500 X 500 $nm^2$ AFM image) of the intermediate (e) and final (f) stages of the etching process for Ga on GaAs(001). Please note the droplet persisting during the whole etching process at the center of the nanohole[76]. Courtesy of David Fuster, Instituto de Micro y Nanotecnología - CNM, CSIC. (g) The emission energy of the GaAs/AlGaAs QD obtained by infilling the nanohole can be controlled by changing GaAs amount of material filling.

**Figure 4:** (a) Photoluminescence spectra of single isolated QDs of GaAs/AlGaAs (top) and InAs/InAlAs (bottom) grown on (111)A substrates. Clear emission from exciton (X), biexciton (XX), and charged trions ($X^+$,$X^-$) is observed in both samples. (b) Magneto photoluminescence of a GaAs QD on (111)A, showing the emergence of optically-forbidden lines for both charged and neutral excitons. The sketch illustrates the recombination selection rule between electrons and mixed heavy holes [158]. (c) Sketch of a Rb-based quantum memory for photons emitted by GaAs

QDs (courtesy of Prof. Eden Figueroa, Stony Brook University): The single photons emitted by an optically excited QD embedded in an optical cavity to increase extraction efficiency, are directed to a Rb vapour cell. Photons, in resonance with a Rb optical transition, are stored by transferring their state to a collective excitation of the Rb cloud by means of control laser pulses. Photon retrieval is also controlled optically and the photon is then fed into a Hanbury Brown and Twiss setup. (d) $g_2(\tau)$ measurements of photons emitted by a biexciton (XX) confined in a LDE GaAs QD under two-photon excitation showing state-of-the-art single-photon purity (Ref.[103]). (e) Hong-Ou-Mandel (HOM) two-photon interference between subsequent photons emitted by a QD showing a high degree of photon indistinguishability without resorting to spectral filtering or Purcell effect. From Ref.[104]. (f) HOM interference between photons emitted by two remote GaAs QDs excited via phonon-assisted two-photon excitation for copolarized (green) and cross-polarized (black) configurations. One of the QDs (QD B in the sketch) is strain-tuned to achieve color coincidence. (g) HOM visibility as a function of emission energy detuning (see color-coded PL spectra of the two QDs in the inset). From Ref. [102].
(h) An LED based on GaAs QDs with emission energy tunable to transition lines of Rb atoms. Left: sketch showing a strain-tunable LED integrated on a piezoelectric substrate launching photons through a Rb cell. Middle: Color-coded electroluminescence (EL) spectra of a QD emission line scanned through the D2 transitions of 87Rb by varying the electric field Fp applied to the strain actuator. Right: EL intensity corresponding to the dashed line in the middle panel. (From Ref. [111]).

**Figure 5**: (a) Anisotropy-induced FSS for [100] and [111] grown GaAs DE QDs. The more symmetric QDs grown on [111] oriented substrate show a lower, on average, and nearly zero FSS, as expected (from Ref.[60]). (b) Correlation between QD FSS value and asymmetry of the nanohole in LDE QDs. A value below resolution of FSS is observed in the case of symmetric nanohole obtained by Al droplet driven etching on AlGaAs (from Ref.[80]). (c) Comparison of FSS of LDE GaAs QDs and InGaAs QDs grown on GaAs(001) substrates (from Ref. [29]). (d) Density matrix for the two-photon state generated by the biexciton-exciton cascade in a LDE GaAs QD with a FSS of 1.2 μeV under two-photon excitation. (e) Left: Fidelity of the two-photon state to the maximally entangled Bell state for 4 GaAs QDs with different FSS and comparison with model calculations. For the QD with smallest FSS, the achieved fidelity is the highest reported so far for polarization-entangled photons generated by QDs. Right: Model calculations of fidelity vs FSS for GaAs and InGaAs QDs with different lifetimes. Different from InGaAs QDs, GaAs QDs have the potential of reaching near unity fidelity because of reduced spin-scattering and short radiative X lifetime (~250 ps). From Ref. 104

Figure 1

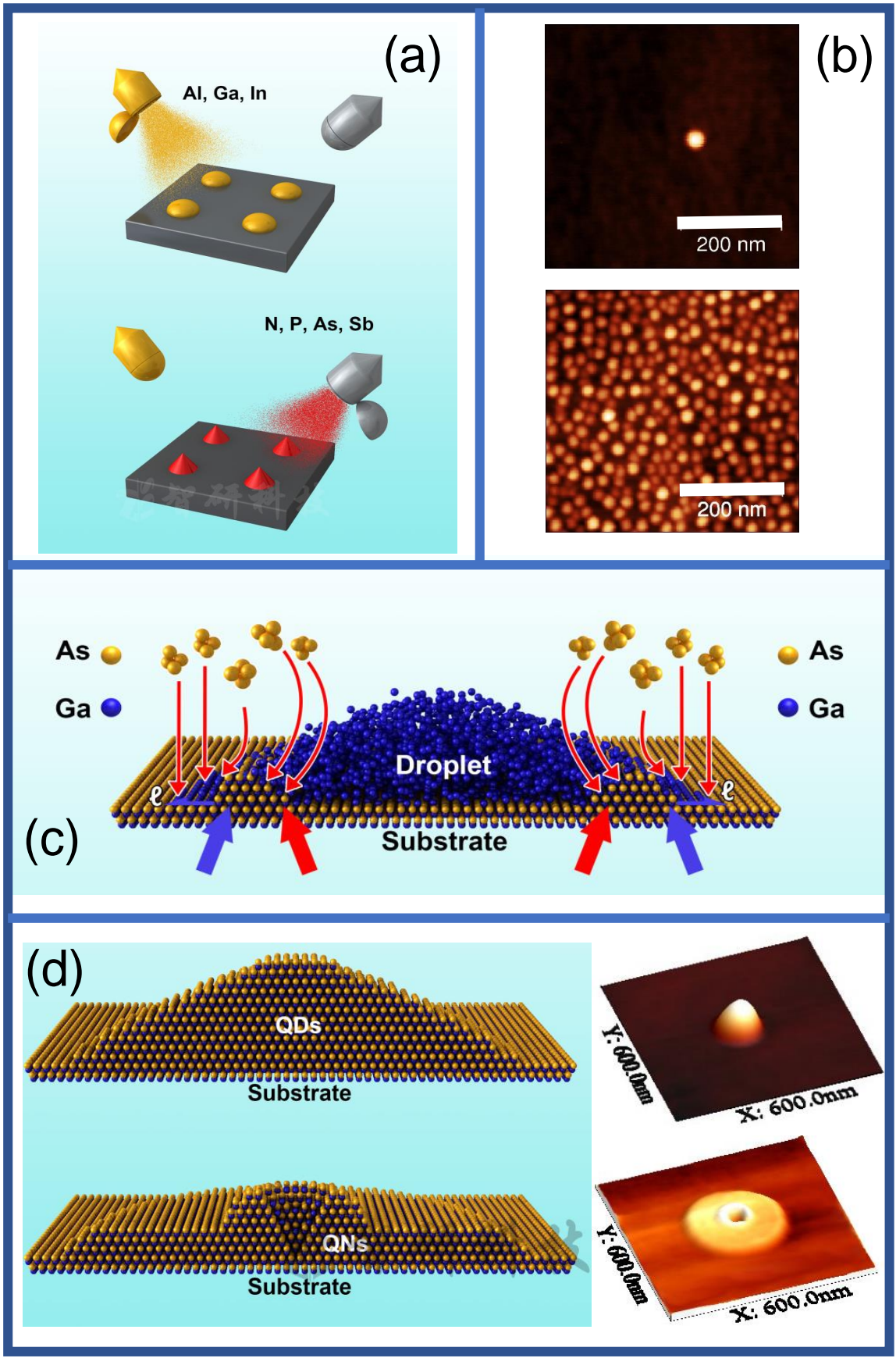

Figure 2

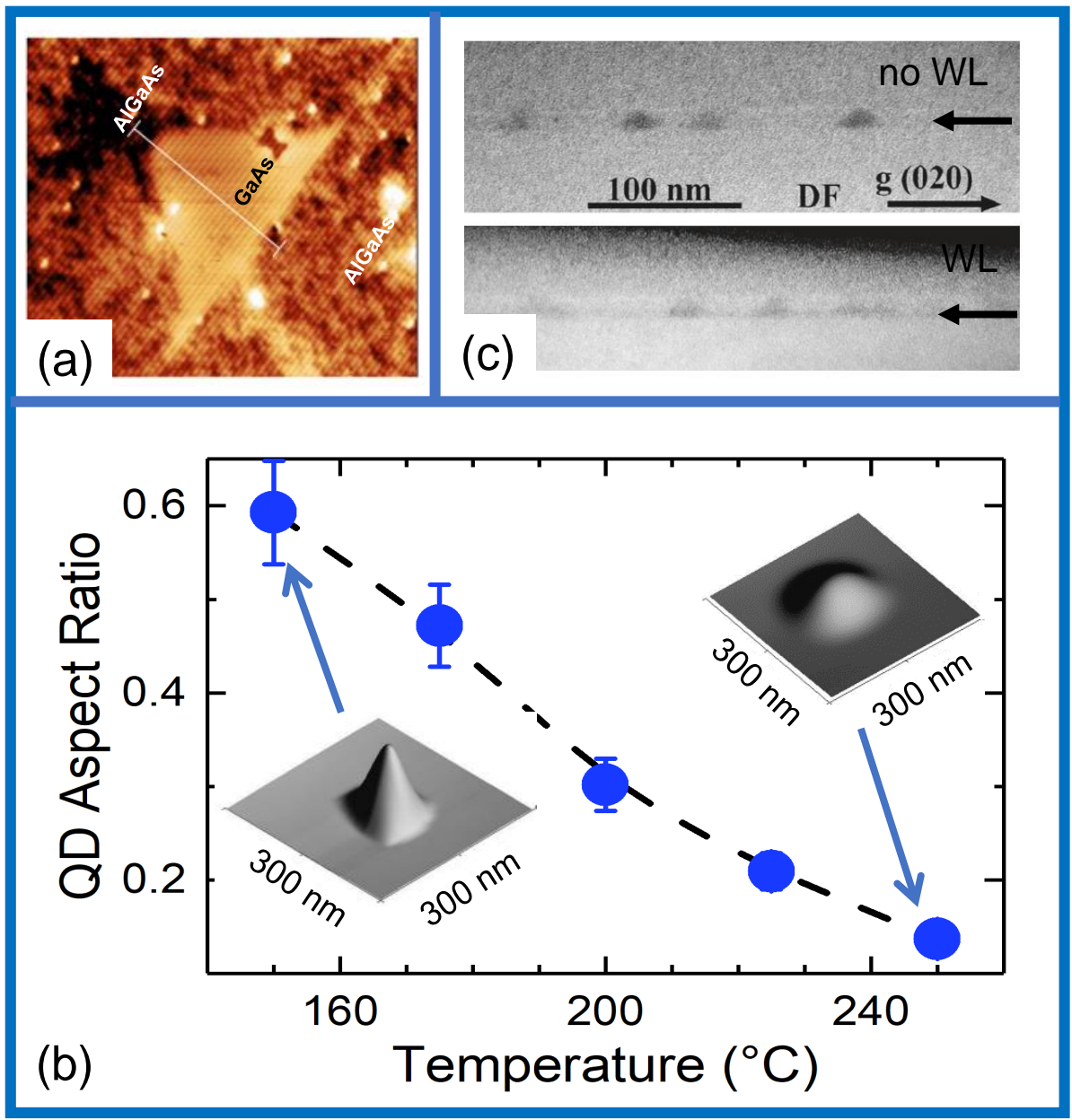

Figure 3

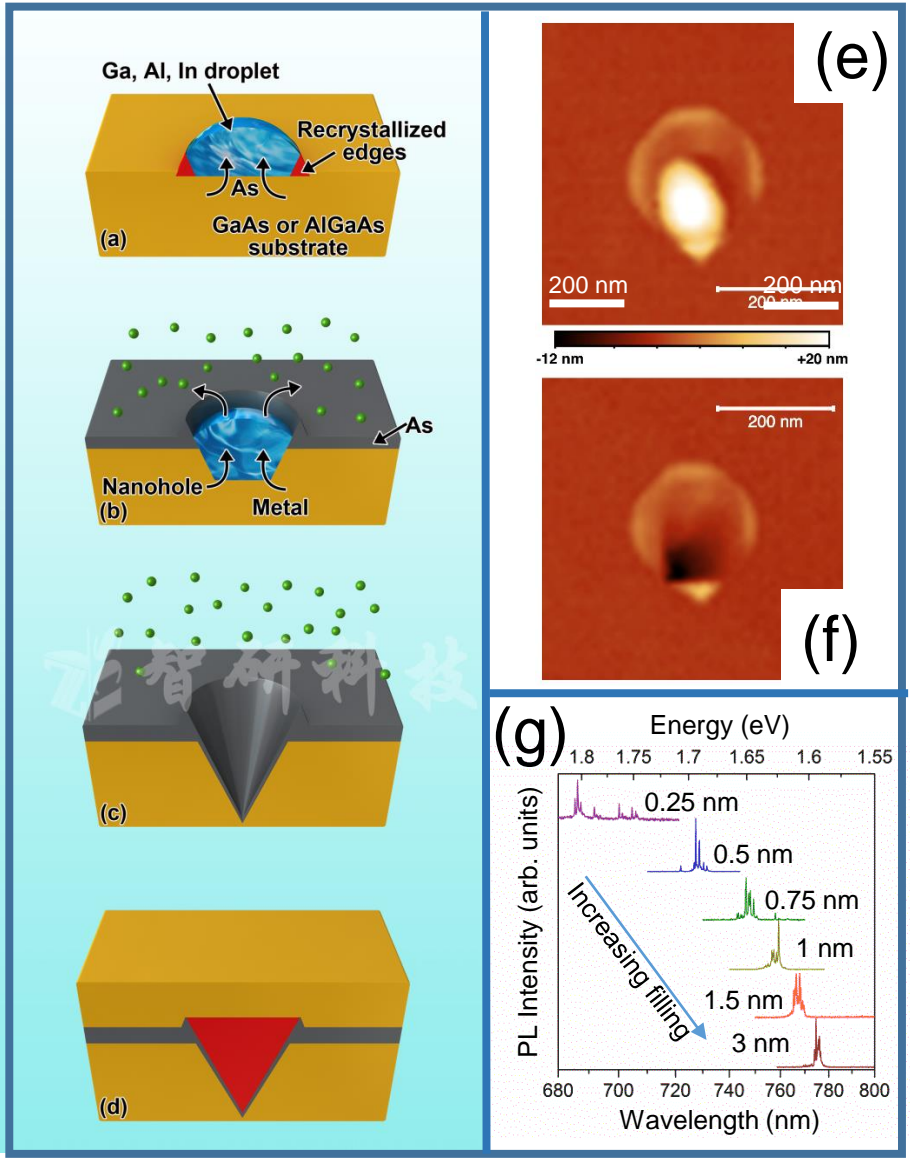

Figure 4

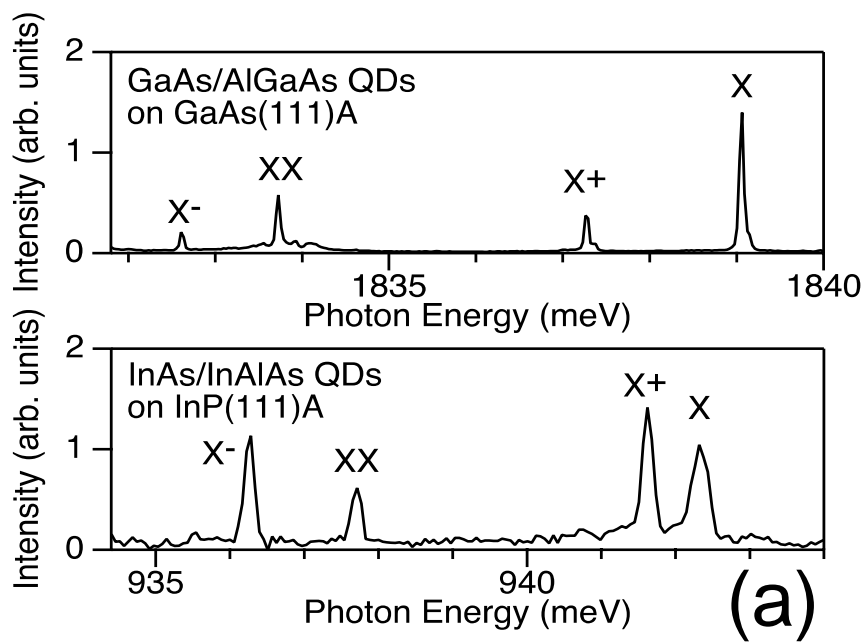
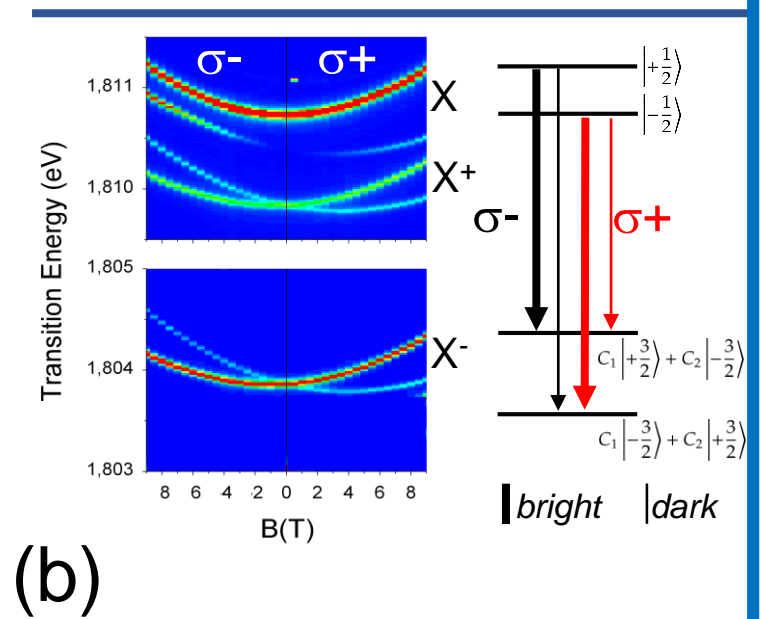
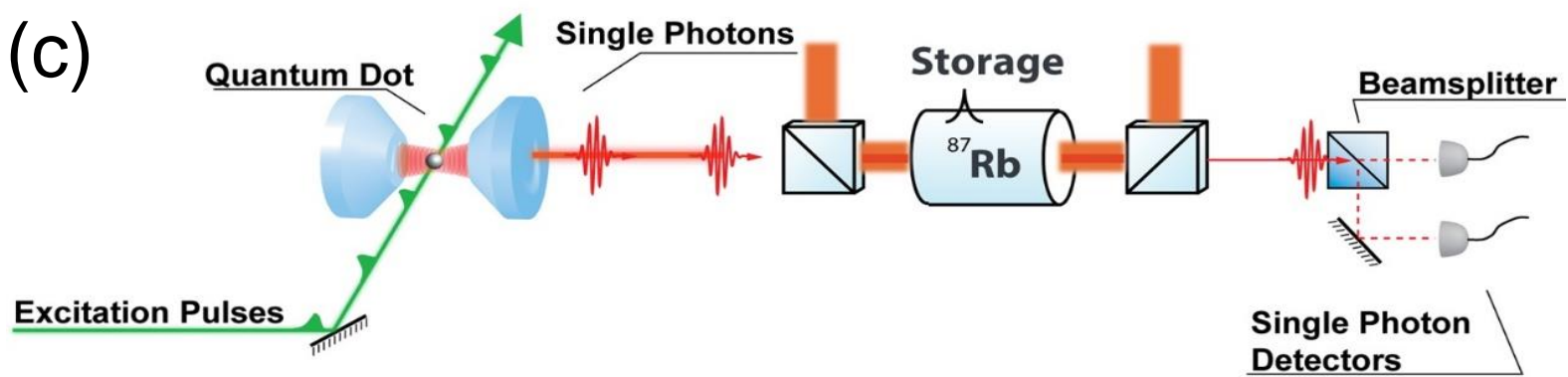
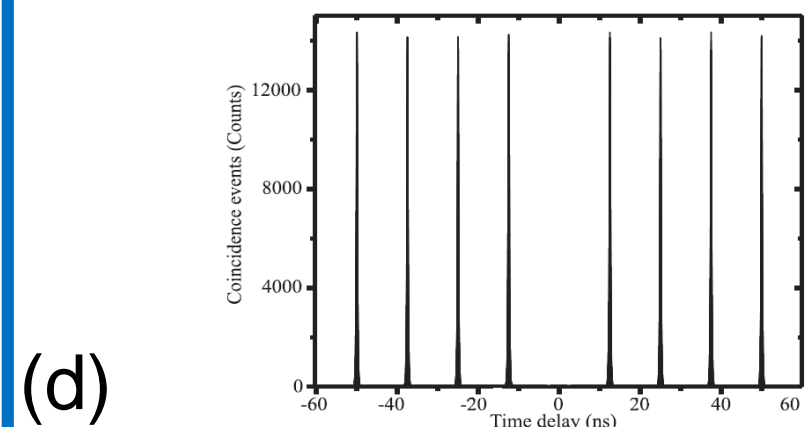
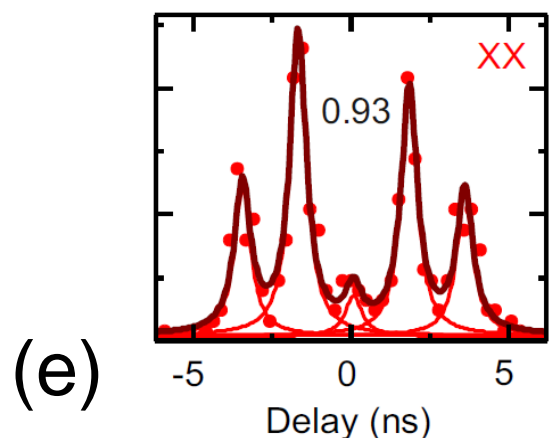
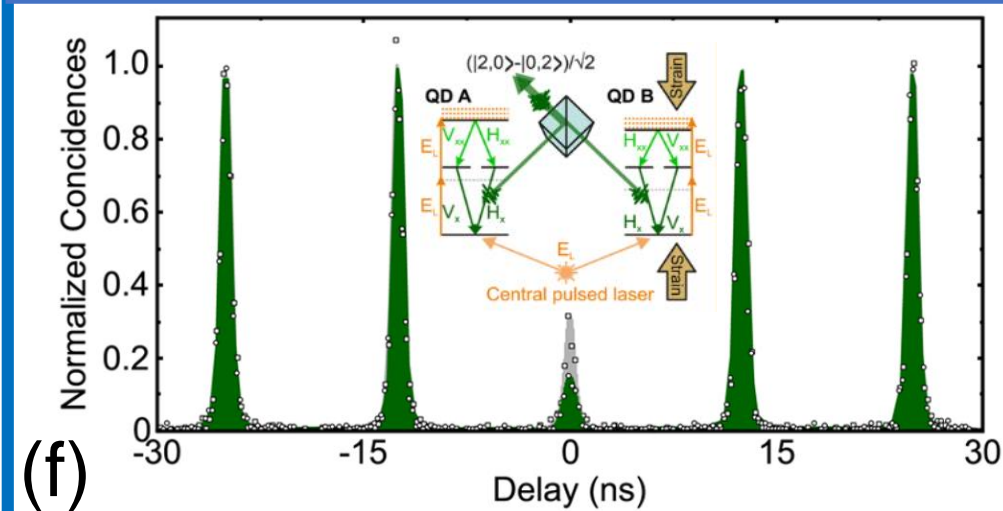
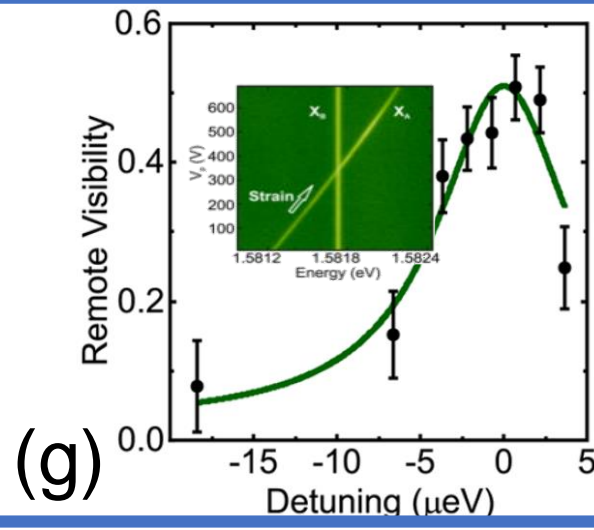
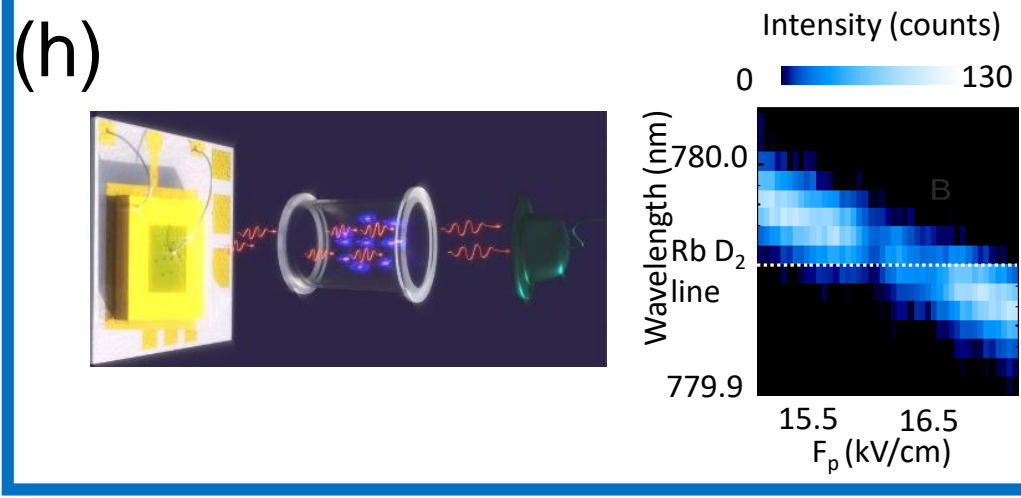
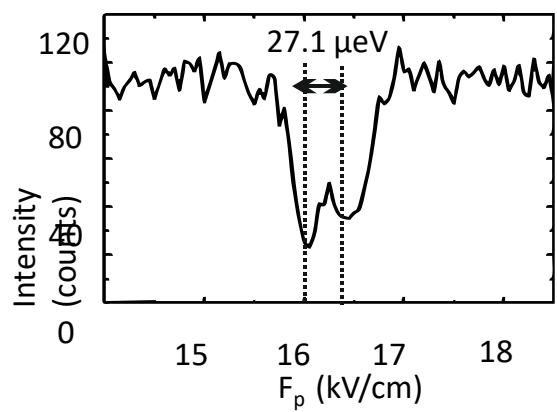

Figure 5

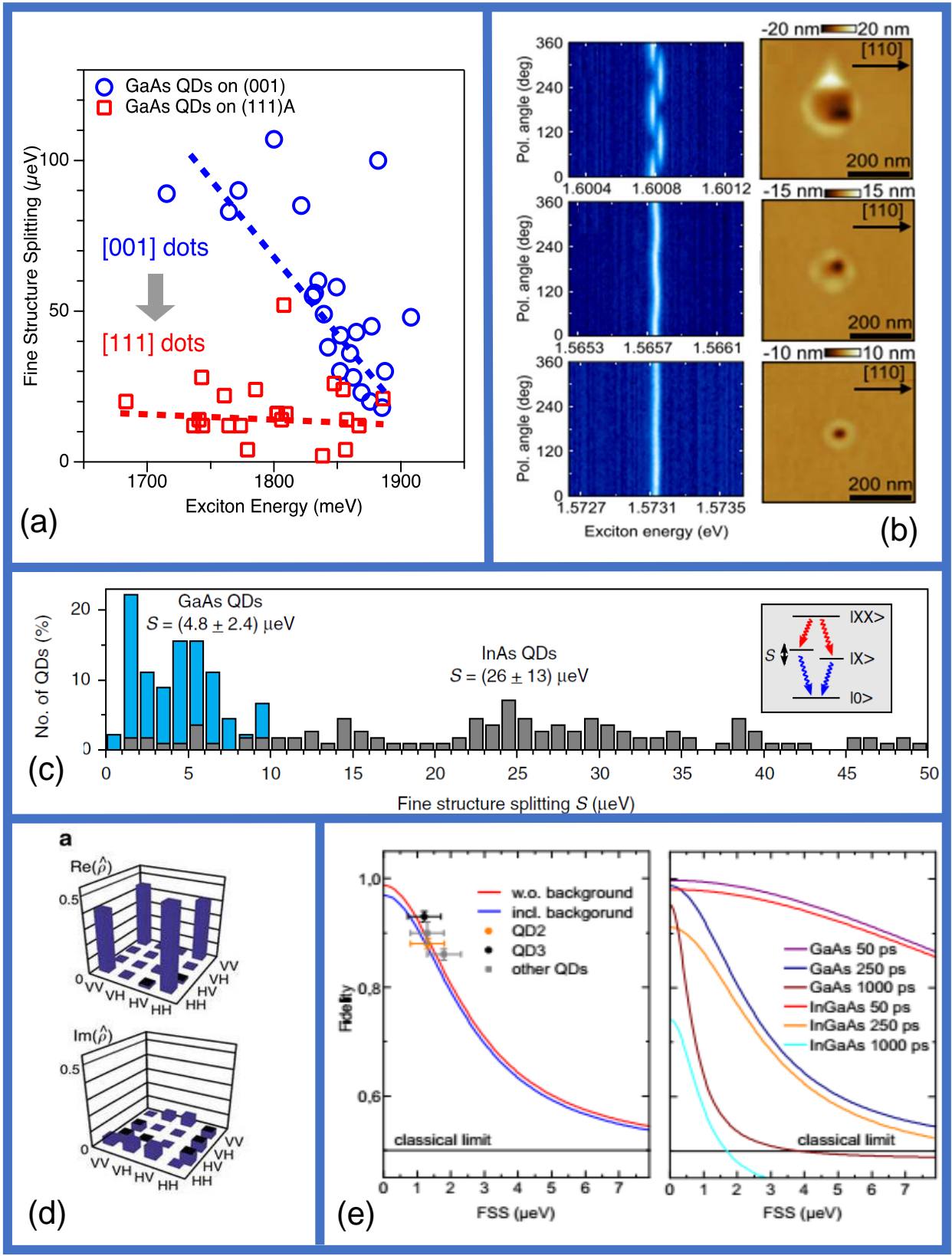